\documentclass[aps,prl,twocolumn,showpacs,floatfix]{revtex4}
\usepackage{graphicx}
\usepackage{dcolumn}
\usepackage{bm}
\usepackage{epsf}
\usepackage{subfigure}
\usepackage{epstopdf}
\usepackage{amsmath}
\usepackage{amssymb}

\newcommand{\mean}[1]{\langle #1 \rangle}

\newcommand{\e}{{\rm e}}

\begin{document}

\title{Three detailed fluctuation theorems}

\author{Massimiliano Esposito}
\affiliation{Center for Nonlinear Phenomena and Complex Systems, Universit\'e Libre de Bruxelles, 
CP 231, Campus Plaine, B-1050 Brussels, Belgium.}
\author{Christian Van den Broeck}
\affiliation{Faculty of Sciences, Hasselt University, B 3590 Diepenbeek, Belgium.}

\date{\today}

\begin{abstract}
The total entropy production of a trajectory can be split into an adiabatic and a non-adiabatic contribution, deriving respectively from the breaking of detailed balance via nonequilibrium boundary conditions or by external driving. We show that each of them, the total, the adiabatic and the non-adiabatic trajectory entropy, separately satisfies a detailed fluctuation theorem.  
\end{abstract}



\maketitle

Small systems are subject to fluctuations. As a result, the energy $Q$ that such a system absorbs from its surrounding reservoir during a given time interval will be a random variable, functional of the specific trajectory followed by the state of the system in the meantime. If the reservoir can be regarded as an idealized thermal reservoir, remaining all the time at equilibrium at a given temperature $\beta^{-1}$ (we set $k_B=1$), one concludes that the change in its entropy $\Delta S_r=-\beta Q$ is also a random variable. Such trajectory entropy has to be distinguished from its ensemble average $\mean{\Delta S_r}$ corresponding to the conventional thermodynamic entropy. In thermostated \cite{Thermostated} and stochastic systems \cite{Stochastic}, it was shown that the probability distribution of $\Delta S_r$ obeys a asymptotic detailed fluctuation theorem (FT) of the form $P(\Delta S_r)/P(-\Delta S_r) \sim \exp(\Delta S_r)$. The result was derived under steady state conditions for a system with bounded energy and in the limit of a large time interval. It was subsequently realized \cite{Seifert05} that this asymptotic FT resulted from a detailed FT for the total entropy $\Delta S_{tot}$, obtained by adding the system entropy $\Delta S$ to the reservoir entropy $\Delta S_r$, valid for all times and for any initial condition, $P(\Delta S_{tot})/\bar{P}(-\Delta S_{tot})= \exp(\Delta S_{tot})$. The overbar tilde refers to the fact that in presence of an external driving, the time-reversed driving has to be considered. It implies an integral FT, namely $\langle\exp(-\Delta S_{tot})\rangle=1$, leading in its turn to $\langle \Delta S_{tot} \rangle \geq 0$, in agreement with the second law of thermodynamics. The asymptotic FT follows in long time limit when $\Delta S$ is bounded and $\Delta S_r \sim t$ becomes dominant. For systems with non-bounded energy the steady state FT can be broken \cite{NotBounded}. In case of an externally driven system initially at (canonical) equilibrium with temperature $\beta^{-1}$, the total entropy production can be expressed as $\Delta S_{tot}=\beta(W-\Delta F)$ where $W$ denotes the work performed on the system and $\Delta F$ is the free energy difference of the system between its final and initial equilibrium states. The detailed FT thus becomes the Crooks relation $P(W)/\bar{P}(-W)= \exp[\beta(W-\Delta F)]$ and its integrated version, the Jarzynski equality $\langle \exp(-\beta W) \rangle = \exp(-\beta \Delta F)$. Both FTs have been derived for stochastic as well as microscopic dynamics \cite{Jarzynski97,Crooks} with the assumption of weak coupling between system and reservoir (see however also \cite{JarzynskiReply04}). Other FTs have been derived. The integral (detailed) Hatano-Sasa FT \cite{Hatano,Harris} (\cite{Chernyak}) for the quantity $\beta Q_{ex}-\Delta S$ is a generalization of the Jarzynski (Crooks) relation to situations where the system starts and ends in a nonequilibrium steady state. $Q_{ex}$ is called the excess heat. We finally mention the Speck-Seifert integral FT for the so-called housekeeping heat $Q_{hk}=Q-Q_{ex}$ \cite{SpeckSeifert05,Harris}. 

The work and FT discussed above basically extend the pioneering ideas of Onsager on the implications of micro-reversibility to (small) nonequilibrium systems. Their application and  experimental verification have been the object of several reviews \cite{Review,Harris,EspositoRMP}. Nevertheless, there is still some confusion in the literature due to the various forms of the FTs and the conditions under which they apply. The purpose of this letter is to clarify the situation with the derivation of three basic FTs. One of them is the ``standard" theorem for the total entropy production $\Delta S_{tot}$. The aforementioned natural separation into a reservoir contribution $\Delta S_r$ and system contribution $\Delta S$ suffers from the deficiency that these quantities themselves do not obey a FT. As we proceed to show there is however another separation of the total entropy, $\Delta S_{tot}=\Delta S_{na}+\Delta S_{a}$, where the so-called non-adiabatic $\Delta S_{na}$ and adiabatic component $\Delta S_{a}$ satisfy, each separately, a detailed and integral FT. These contributions have a physical meaning as they correspond to two basic ways in which a system can be brought out of equilibrium, namely by driving the system by a time dependent force or by applying steady nonequilibrium constraints. Furthermore, in the case of a single heat reservoir $\Delta S_{a}=-\beta Q_{hk}$ and if in addition only transitions between steady states are considered $\Delta S_{na}=\Delta S-\beta Q_{ex}$ \cite{Esposito07}. Under these conditions, the aforementioned corresponding FTs are in fact  FTs for $\Delta S_{a}$ and $\Delta S_{na}$.

Before embarking on the general derivation of the FTs for total entropy, and its adiabatic and non-adiabatic contributions, we make a number of revealing mathematical statements which will clarify and simplify the subsequent discussion, and which streamlines the derivation of FTs. A random variable $R$, with probability distribution $P(R)$, is said to obey an integral FT if:
\begin{equation}\label{ift}
\langle e^{-R} \rangle = \int dR P(R) e^{- R}=1.
\end{equation}
By Jensen's inequality, we then also find that $\langle R \rangle \ge 0$ [average with respect to $P(R)$]. An alternative formulation of (\ref{ift}) is to state that the positive quantity $\tilde{P}(R) =P(-R)e^{R}$ is a normalized probability distribution. We conclude that when a variable obeys an integral FT, it automatically obeys a detailed FT, i.e., there exists a probability distribution $\tilde{P}$ of $R$ such that:
\begin{equation}\label{dft}
\frac{P(R)}{\tilde{P}(-R)} = e^{ R} \;\;\;\;\; \forall R.
\end{equation}
The previously discussed entropy-related FTs are of the above form with $R$ playing the role of the total trajectory entropy production, and with the additional feature that the distribution $\tilde{P}(R)$ is simply related to the original distribution $P(R)$ as $\tilde{P}(R)={P}(R)$ or $\tilde{P}(R)=\bar{P}(R)$.

We next present a recipe to generate a variable obeying a FT. We make the observation that the relative entropy or Kullback-Leibler distance between two probability distributions ${\cal P}_{\bold m}$ and ${\tilde{\cal P}}_{\bold m}$ of a (set of) random variable(s) ${\bold m}$ is:
\begin{equation}\label{re}
D({\cal P}||{\tilde{\cal P}})=\sum_{\bold m}  {{\cal P}_{\bold m}} \ln \frac{{\cal P}_{\bold m}}{{\tilde{\cal P}}_{\bold m}} \ge 0.
\end{equation}
The variables $\bold m$ correspond here to the trajectories followed by the system during a given time interval. ${\cal P}$ and ${\tilde{\cal P}}$ denote normalized probability densities of these trajectories. Performing the sum over ${\bold m}$ is used as a compact notation for a path integral over these trajectories. For a single realization of a trajectory, the quantity of interest is thus the variable $r_{\bold m}$:
\begin{equation}\label{rQTY}
r_{\bold m} = \ln\frac{{\cal P}_{\bold m}}{{\tilde{\cal P}}_{\bold m}}.
\end{equation}
We now consider the probability distribution of this variable, when ${\bold m}$ is sampled according to ${\cal P}_{\bold m}$. The resulting random variable $R$ is characterized by the probability distribution 
\begin{eqnarray}
P(R) = \sum_{\bold m} \delta (R-r_{\bold m}) {\cal P}_{\bold m}.  \label{dftaver}
\end{eqnarray}
One verifies that it obeys the FT (\ref{ift}): 
\begin{eqnarray}
\int dR P(R) e^{- R}=\sum_{\bold m} {\cal P}_{\bold m} \e^{-r_{\bold m}}= \sum_{\bold m} {\tilde{\cal P}}_{\bold m}=1. \label{TildeDistrib}
\end{eqnarray}
We now assume that ${\tilde{\cal P}}_{\bold m}$ can be obtained via a mathematical ``tilde" operation from ${\cal P}_{\bold m}$ which is an involution:
\begin{eqnarray}
\tilde{\tilde{\cal P}}_{\bold m}={\cal P}_{\bold m} \label{InvCond}.
\end{eqnarray}
As a result, we find that 
\begin{equation}\label{tilderQTY}
{\tilde{r}}_{\bold m} = \ln\frac{{\tilde{\cal P}}_{\bold m}}{{\cal P}_{\bold m}}=-r_{\bold m},
\end{equation}
which implies that the probability distribution $\tilde{P}(R)$ can be written as follows:
\begin{eqnarray}
\tilde{P}(R) &=& P(-R) \e^{R} 
= \sum_{\bold m} \delta (R+r_{\bold m}) {{\cal P}}_{\bold m}  \e^{-r_{\bold m}} \nonumber \\
&=& \sum_{\bold m} \delta (R-{\tilde{r}}_{\bold m}) \tilde{{\cal P}}_{\bold m} \label{tildePR}
\end{eqnarray}
and justifies a-posteriori the use of the same superscript tilde for both distributions $\tilde{P}(R)$ and $\tilde{{\cal P}}_{\bold m}$. 

To apply the above mathematical observations to the problem of entropy production in a driven nonequilibrium physical system, we need to introduce a physical model for the dynamics of the system. For simplicity and clarity of presentation, we restrict ourselves here to the case of Markovian dynamics on a set of discrete states $m$. The transition probability per unit time from state $m'$ to $m$ via a mechanism $\nu$ will be denoted by $W_{m,m'}^{(\nu)}(\lambda_{t})$. The latter generally depend on time, via a control variable $\lambda_{t}$ which describes external driving. After introducing the compact notation $W_{m,m'}(\lambda_{t})= \sum_{\nu} W_{m,m'}^{(\nu)}(\lambda_{t})$, the probability ${p}_m(t)$ to be in state $m$ at time $t$ obeys the following Master equation:
\begin{eqnarray}
\dot{p}_m(t) = \sum_{m'} W_{m,m'}(\lambda_{t}) p_{m'}(t).
\label{OrigME}
\end{eqnarray} 
Consistent with the notation used earlier, we introduce the probability  ${\cal P}_{\bold m}$ for a trajectory $\bold m=\{m_t, t\in[0,T]\}$, obeying the Markovian stochastic dynamics (\ref{OrigME}). Note that ${\cal P}_{\bold m}$ depends implicitly on the applied schedule ${\bold \lambda}=\{\lambda_{t}, t\in[0,T]\}$ of the control variable and on the probability $p_{m_0}(0)$ of its initial state ${m_0}$. To obtain an explicit expression for ${\cal P}_{\bold m}$, we note that a trajectory is  described by its initial condition ${m_0}$, the times $\tau_{j}$ at which is undergoes  changes in its state from $m_{j-1}$ (state prior to jump) to $m_{j}$ (state after  jump) and the type $\nu_{j}$ of this transition, with the index $j$ running over the total number $N$ of jumps of the specified trajectory. We use $\tau_{0}=0$ and $\tau_{N+1}=T$. We can thus write:
\begin{eqnarray} \label{ExplProb}
{\cal P}_{\bold m}&=& p_{m_0}(0) \big[ \prod_{j=1}^{N} \e^{\int_{\tau_{j-1}}^{\tau_{j}} d\tau' 
W_{m_{j-1},m_{j-1}}(\lambda_{\tau'})} \\ &&\hspace{0cm} \times W_{m_{j},m_{j-1}}^{(\nu_{j})}(\lambda_{\tau_j}) \big] 
\e^{\int_{\tau_{N}}^{T} d\tau' W_{m_{N},m_{N}}(\lambda_{\tau'})} \;.\nonumber
\end{eqnarray}
In the sequel, we will again use an overbar for the time-reversed quantities: $\bar{\bold m}$ stands for the time-reversed trajectory of ${\bold m}$,  $\bar{\bold \lambda}$ for time reversed control schedule of ${\bold \lambda}$ (i.e.  $\bar{\lambda}_t=\lambda_{T-t}$), and $\bar{\cal P}_{\bar{\bold m}}$ for the probability of the time-reversed trajectory  $\bar{\bold m}$ (the initial condition of $\bar{\bold m}$ being obviously the final state ${m_N}$ of ${\bold m}$, appearing with probability $p_{m_N}(T)$), with time-reversed driving $\bar{\bold \lambda}$. Explicitly:
\begin{eqnarray}
\bar{\cal P}_{\bar{\bold m}} &=& \e^{\int_{T-\tau_{1}}^{T} d\tau' W_{m_{0},m_{0}}(\bar{\lambda}_{\tau'})} 
\big[ \prod_{j=1}^{N}  W_{m_{j-1},m_{j}}^{(\nu_{j})}({\bar{\lambda}}_{T-\tau_{j}}) \nonumber \\ &&\hspace{0cm} 
\times \e^{ \int_{T-\tau_{j+1}}^{T-\tau_{j}} d\tau' W_{m_{j},m_{j}}(\bar{\lambda}_{\tau'}) } \big]  p_{m_N}(T) \;.  \label{ExplProbBack}
\end{eqnarray}

To make the connection with entropy production, we start from the following expression for the total trajectory entropy production, obtained from both stochastic and microscopic analysis \cite{Seifert05,VandenBroeck07,Esposito07,Review,Harris}:
\begin{eqnarray}
\label{tet}
\Delta S_{tot}[{\bold m}] = \ln \frac{{\cal P}_{\bold m}}{\bar{\cal P}_{\bar{\bold m}}} .
\end{eqnarray}
Inserting the above explicit formula of ${\cal P}_{\bold m}$ in terms of the trajectories, one finds that, upon taking the ratio ${{\cal P}_{\bold m}}/{\bar{\cal P}_{\bar{\bold m}}}$, only the contributions of the jumps in the state of the system survive, together with a contribution in the change of probability of initial versus final state:
\begin{eqnarray}
\Delta S_{tot}[{\bold m}] = \ln \frac{p_{m_0}(0)}{p_{m_N}(T)} +
\sum_{j=1}^{N} \ln \frac{W_{m_{j},m_{j-1}}^{(\nu_{j})}(\lambda_{\tau_j})}
{W_{m_{j-1},m_{j}}^{(\nu_{j})}(\lambda_{\tau_{j}})} . \label{caaae}
\end{eqnarray}
The total trajectory entropy is thus given as a sum, $\Delta  S_{tot}[{\bold m}]=\Delta S[{\bold m}]+\Delta S_r[{\bold m}]$, with the first term in the r.h.s., $\Delta S[{\bold m}]= -\ln{p_{m_N}(T)}-(-\ln {p_{m_0}(0)})$, corresponding to the change in trajectory system entropy, and the second, $\Delta S_r[{\bold m}]$, being the change in trajectory reservoir entropy along the specified trajectory. 
 
To proceed, we introduce the  instantaneous stationary distribution $p^{st}_m(\lambda_t)$, which is the steady state that is reached when the transition probabilities are frozen in time to the value $W_{m,m'}(\lambda_{t})$. They correspond to the normalized zero right eigenvector of this transition matrix:
\begin{eqnarray}
\sum_{m'} W_{m,m'}(\lambda_{t}) p^{st}_{m'}(\lambda_t)=0.
\end{eqnarray}
The total trajectory entropy (\ref{caaae}) can now be split in the following alternative way:
\begin{eqnarray}
\Delta S_{tot}[{\bold m}] = \Delta S_{na}[{\bold m}] + \Delta S_{a}[{\bold m}]\label{osplit}
\end{eqnarray}
with the non-adiabatic trajectory entropy:
\begin{eqnarray} \label{TEPna}
\Delta S_{na}[{\bold m}] = \ln \frac{p_{m_0}(0)}{p_{m_N}(T)} +
\sum_{j=1}^{N} \ln \frac{p_{m_{j}}^{\rm st}(\lambda_{\tau_j})}{p_{m_{j-1}}^{\rm st}(\lambda_{\tau_j})} ,
\end{eqnarray} 
and the adiabatic contribution:
\begin{eqnarray} \label{TEPa}
\Delta S_{a}[{\bold m}] = \sum_{j=1}^{N} \ln \frac{ W_{m_{j},m_{j-1}}^{(\nu_{j})}(\lambda_{\tau_j}) 
p_{m_{j-1}}^{\rm st}(\lambda_{\tau_j}) }{ W_{m_{j-1},m_{j}}^{(\nu_{j})}(\lambda_{\tau_{j}}) p_{m_{j}}^{\rm st}(\lambda_{\tau_j}) } .
\end{eqnarray} 
We first note that this separation is relevant if the system steady state does not satisfy detailed balance, else the argument of the logarithm in (\ref{TEPa}) equal one and $\Delta S_{a}=0$. The name given to these two contributions can be justified as follows. Suppose that the relaxation of the stochastic dynamics is very fast compared to the timescale of the schedule $\lambda_t$. During such a so-called adiabatic process, the probability distribution will assume at all times the instantaneous steady state form $p^{st}_m(\lambda_t)$ and it can be verified that this implies $\mean{\Delta S_{na}}=0$ and thus $\mean{\Delta S_{tot}}=\mean{\Delta S_{a}}$. When such a timescale separation does not exist, $\mean{\Delta S_{na}} \neq 0$ and it is  therefore natural to refer to it as the non-adiabatic contribution.

The total trajectory entropy (\ref{tet}) is the logarithm of the ratio of trajectory probabilities. We now show that the adiabatic and non-adiabatic entropy productions can also be written under such a form by introducing the Markov process that is obtained from the original one, with transition matrix $W^{(\nu)}(\lambda_{t})$, by considering the adjoint  transition matrix $W^{(\nu)\; +}(\lambda_{t})$ (also called dual or reversal \cite{Crooks,Chernyak}) given by: 
\begin{eqnarray}
W^{(\nu)\; +}_{m,m'}(\lambda_{t}) =\frac{W_{m',m}^{(\nu)}(\lambda_{t}) p_m^{st}(\lambda_t)}{p_{m'}^{st}(\lambda_t)}.
\label{ae}
\end{eqnarray}
This Markov process has the same instantaneous steady states as the original one. $W^+$ coincides with $W$ only under the condition that detailed balance is satisfied with respect to the instantaneous stationary distribution $p^{st}_m(\lambda_t)$. We will use the notation $\cal P^+_{\bold m}$ for the probabilities of trajectories that are generated by the Markov processes associated to $W^+(\lambda_{t})$ and $\bar{\cal P}^+_{\bar{\bold m}}$ denotes the probabilities of time reversed trajectories generated with the adjoint rate matrix with time reversed driving, i.e. $W^+(\lambda_{T-t})$. 

The main point in the derivation of the FTs is to note that the non-adiabatic and adiabatic entropy production contributions are again given in terms of the logarithm of the ratio of trajectory probabilities: 
\begin{eqnarray}
\Delta S_{na}[{\bold m}] \equiv \ln \frac{{\cal P}_{\bold m}}{{\bar{\cal P}}^{+}_{\bar{\bold m}}}  \  \ , \ \
\Delta S_{a}[{\bold m}] \equiv \ln \frac{{\cal P}_{\bold m}}{{\cal P}^+_{\bold m}} \label{rnara}
\end{eqnarray}
This can be verified by inspection, using (\ref{ExplProb}) and (\ref{ExplProbBack}) with the appropriate transition probabilities for the trajectory probabilities. In particular the diagonal elements of $W^ +$ and $W$ are identical, so that again only contributions from jumps and initial conditions appear. The expression for the non-adiabatic trajectory entropy change $\Delta S_{na}$ is similar to that for total trajectory entropy $\Delta S_{tot}$, since they both feature the ratio of a forward and time-reversed path. In both cases, the time-reversed path starts with the final probability distribution of the forward evolution, but determined respectively by the (time-reversed) transition matrix $W(\bar{\bold \lambda})$ and by the adjoint (time-reversed) transition matrix $W^+(\bar{\bold \lambda})$. The adiabatic trajectory entropy change $\Delta S_{a}$ on the other hand is given in terms of  the ratio of trajectory probabilities with same initial condition and same driving, but described by the ``forward" transition matrix $W$ and the adjoint transition matrix $W^+$, respectively. It vanishes when $W^+=W$.

The trajectory entropies (\ref{tet}) and (\ref{rnara}) have the structure of the random variable $r_{\bold m}$, cf. (\ref{rQTY}). Hence they satisfy a (detailed and integral) FT when they are sampled with the ``forward" probability ${\cal P}_{\bold m}$.  We furthermore identify $\tilde{P}_{\bold m}={{\bar{\cal P}}_{\bar{\bold m}}}$ for the total trajectory entropy production, $\tilde{P}_{\bold m}={{\bar{\cal P}}^{+}_{\bar{\bold m}}}$ for its non-adiabatic contribution and $\tilde{P}_{\bold m}={{{\cal P}}^{+}_{{\bold m}}}$ for its adiabatic one. It is obvious that the involution condition (\ref{InvCond}) is satisfied for all of them. From (\ref{tilderQTY}), we thus find that $\tilde{r}_{\bold m}$ is given by $-\Delta S_{tot}[\bold m]$, $-\Delta S_{na}[\bold m]$ and $-\Delta S_{a}[\bold m]$ for total, non-adiabatic and adiabatic trajectory entropy, respectively. We conclude from (\ref{tildePR}) and  (\ref{dft}) that the  following three detailed FT hold: 
\begin{eqnarray}
&&\hspace{0.6cm}\frac{P(\Delta S_{tot})}{{\bar{P}}(-\Delta S_{tot})} = e^{\Delta S_{tot}}  \label{FT1} \\
&&\hspace{-1.3cm} \frac{P(\Delta S_{na})}{{\bar{P}}^+(-\Delta S_{na})} = e^{\Delta S_{na}} \ \ \;, \ \ 
\frac{P(\Delta S_{a})}{{P}^+(-\Delta S_{a})} = e^{\Delta S_{a}} \label{FT3}.
\end{eqnarray}
As a by-product, we also note that each average entropy can be expressed as a relative entropy of the form (\ref{re}), namely
$\mean{\Delta S_{tot}} =D({{\cal P}_{\bold m}}||{\bar{\cal P}_{\bar{\bold m}}}) \ge 0$, $\mean{\Delta S_{na}} = D( {{\cal P}_{\bold m}}||{\bar{\cal P}^+_{\bar{\bold m}}} ) \ge 0$, and $\mean{\Delta S_{a}} = D( {{\cal P}_{\bold m}}||{{\cal P}^+_{\bold m}}) \ge 0$. This identification for the total entropy production has been done for both stochastic \cite{RelEnt} and microscopic dynamics \cite{VandenBroeck07}. 

We close with additional comments. First, we mentioned that the FTs are a consequence of micro-reversibility (and Liouville's theorem). This is not immediately apparent in the above derivation. However, these fundamental properties of micro-dynamics are needed to identify (\ref{tet}) as the correct expression for the total trajectory entropy \cite{gomez2008}. In particular, equilibrium corresponds to the absence of an arrow of time, ${\cal P}_{\bold m}=\bar{\cal P}_{\bar{\bold m}}$, and is thus equivalent with $\Delta S_{tot}[{\bold m}]=0$. Second, the above derived FTs also apply to the case of (driven) Langevin or Fokker-Planck dynamics, since these can be obtained in the appropriate limit from Master Equation dynamics. This procedure allows to make the connection with related results obtained at the level of the Langevin equation \cite{Hatano,SpeckSeifert05,Chernyak}. The results can also be extended to the case of particle transport between system and reservoirs. Third, the experimental measurement of the adiabatic and non-adiabatic trajectory entropy should not pose any problems, since ratios of trajectory probabilities have been measured before \cite{Ciliberto}. Finally, the fact that two constitutive parts of the entropy production have separately properties identical to that of the total entropy, suggests that implications of the fluctuation theorem (e.g., Onsager symmetry) and of the second law itself (e.g., predictions on efficiency of thermal machines), can be unravelled into an adiabatic and non-adiabatic component.
   
\section*{Acknowledgments}

M. E. is supported by the Belgian Federal Government (IAP project ``NOSY").


\end{document}